\documentclass[aps,pre,twocolumn,amsmath,amssymb]{revtex4} 
\usepackage{natbib}       
\usepackage{amsfonts}
\usepackage{graphicx}     
\usepackage{color}
\newcommand{\Eq}[1]{Eq.~(\ref{#1})}
\newcommand{\fig}[1]{Fig.~\ref{#1}}

\begin{document}
\title{Monte Carlo simulations reveal the straightening up of an end-grafted flexible
  chain with a rigid side chain}
\author{Marcel Hellmann$^{1,2}$, Matthias Weiss$^2$, and Dieter W. Heermann$^1$}
\affiliation{$^1$ Institut f\"ur Theoretische Physik, Philosophenweg 19, Universit\"at
  Heidelberg, D-69120 Heidelberg, Germany}
\affiliation{$^2$ Cellular Biophysics Group, German Cancer Research
	Center, Im Neuenheimer Feld 580, D-69120 Heidelberg, Germany}

\begin{abstract}
We have studied the conformational properties of a flexible end-grafted chain 
(length $N$) with a rigid side chain (length $S$) by means of Monte Carlo 
simulations. Depending on the lengths $N$ and $S$ and the branching site, $b$, 
we observe a considerable straightening of the flexible backbone as quantified 
via the gyration tensor. For $b=N$, i.e. when attaching the side chain to the 
free end of the flexible backbone, the effect was strongest.
\end{abstract}

\maketitle

\section{Introduction}
Polymers anchored on a locally flat substrate are of great importance 
for functionalized surfaces in the material sciences~\cite{Brittain2000} 
and in biology~\cite{Sackmann2000}. A particular example for the latter 
is the protective extracellular matrix of living cells that is composed 
of the flexible polymer hyaluronic acid (HA) to which semi-flexible 
aggrecan chains are attached via linker proteins~\cite{LJJ+93, KCW+02}. 
The HA-aggrecan has recently received an increased attention as it is 
not only used to equip cells with a protective layer but also plays an 
important role as gliding surface in the articular cartilage of synovial
joints~\cite{KCW+02}. Understanding the material properties of the 
HA-aggrecan system on the nano- and mesoscale are thus of key importance 
when aiming at constructing biomimetic surfaces that mimic the function 
of natural cartilages. In fact, on the single-molecule scale the HA-aggrecan 
system can be simplified to a flexible polymer, attached to a planar
substrate, with rigid side chains.

Flexible, self-avoiding end-grafted polymers without side chains have been 
studied extensively by theory, experiment and computational 
approaches~\cite{Milner1991,Halperin1991}.
Typically, a large set of end-grafted polymers has been considered, 
where the grafting density~$\sigma$, i.e. the distance between individual 
chains, was varied ('polymer brush'). In the limit of very low $\sigma$, 
the polymer chains can be considered as isolated entities, each acquiring 
approximately a half-spherical shape ('mushroom') with a radius comparable 
to the Flory-radius $R_F = a \cdot N^{3/5}$ of a coil in a good 
solvent~\cite{Gennes1980}. Here, $a$ denotes the size of a monomer. In 
fact, a more detailed description of the polymer shape requires the 
gyration tensor~\cite{Solc1971_1, Solc1971_2}:
\begin{equation}\label{gyrtensor}
  S_{mn} = \frac{1}{N} \sum \limits_{i=1}^{N} r_{m}^{(i)} r_{n}^{(i)},
\end{equation}
where $\mathbf r^{(i)}$ is the position of the $i$th monomer and indices $m,\,n$
denote its individual components. Using an orthogonal transformation, $\mathbf{S}$
can be converted to diagonal form with entries $L_1^2\le L_2^2\le L_3^2$ denoting
the squared lengths of the principal axes of gyration while the trace of
$\mathbf S$ yields the squared radius of gyration. The associated eigenvectors
of the gyration ellipsoid are denoted by $\mathbf v_1$, $\mathbf v_2$, and 
$\mathbf v_3$, respectively.

The ratios of the principle axes of inertia quantify the deviation of the polymer 
shape from a sphere. Using Monte Carlo simulations it was shown that simple 
random walks (RW) reveal a pronounced asphericity ('asphericity of the RW')
mirrored in the asymptotic ratios  
$\langle L_1^2 \rangle:\langle L_2^2 \rangle:\langle L_3^2 \rangle \to 1:2.7:12.0$~\cite{Bruns1992}. 
Self-avoiding walks (SAW) show an even more aspheric shape
($1:2.98:14.0$)~\cite{Bruns1992} that becomes more pronounced when attaching 
the SAW with one end to a flat, solid substrate ($1:3.0:14.9$ for $N \to \infty$)~\cite{Huang2001}.

The conformational properties of polymer systems comprising side chains branching out 
from a flexible or semi-flexible backbone at varying density ('comb polymers') have 
also been considered extensively in theory and simulation studies~\cite{Freire1999}.
The flexibility of the entire complex has been shown to be strongly influenced
by the density of flexible side chains $\sigma_{\ell}$ with three scaling regimes 
in case for a free (i.e. not end-grafted) backbone~\cite{Fredrickson1993, Rouault1996}. 
One of the main results was that attaching a large number of flexible side chains 
stiffens an otherwise flexible backbone. Rigid side chains induced larger local
fluctuations of the backbone as compared to flexible side chains~\cite{Saariaho1999}, 
yet the persistence length of the entire complex was shown to grow super-linear with 
the length of the side chains in contrast to the much weaker dependency in case of 
flexible side chains. In other words, attaching a large number of rigid side chains 
yields an efficient way to stiffen a flexible backbone. 

Here, we investigate the conformation of a flexible, end-grafted self-avoiding chain 
of length $N$ with a {\em single} rigid side chain (length $S$) by means of Monte Carlo 
simulations~\cite{Heermann2002}. We find that attaching the side chain leads to a 
considerable anisotropic swelling of the flexible backbone and a straightening to a more 
brush-like configuration. This phenomenon not only depends on $N$ and $S$, but also on 
the position $b$ at which the side chain is attached. In particular, we show that for 
fixed $N$ and $b$ the squared radius of gyration $R_g^2$ rises sigmoidally from the 
unperturbed value $R_0^2$ with increasing~$S$ and levels off to a constant for 
$S\gg b/2$. In fact, the side-chain dependent swelling of the backbone follows a 
scaling relation with maximum value $\Delta\sim R_0^2b^3/N^2$. Contrary to an undisturbed 
end-grafted chain, the ratio $R_{\rm end}^2/R_g^2$ increases with $S$, thus 
highlighting the successive breaking of isotropy, that is also mirrored in dramatic 
changes of the ratios of the principle axes of gyration. Furthermore, for 
$b\approx N$ and $S\gg b/2$, the distribution of normalized angles 
$\varphi$ ($\langle\varphi\rangle=1$) between the surface normal and the
longest axis of the backbone's gyration ellipsoid follows a Weibull
distribution for all combinations $N,\,b,\,S$.
%
\section{Simulation Method}
\begin{figure} 
\begin{center}
  \includegraphics[width=7cm]{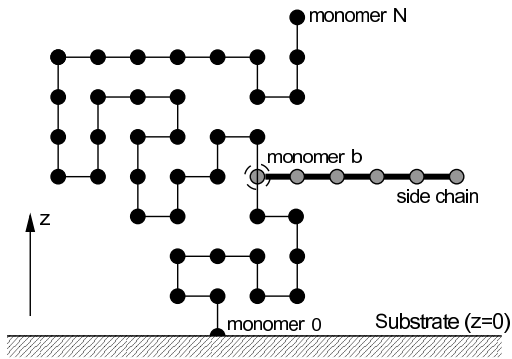}
  \caption{Two-dimensional sketch of the self-avoiding polymer on a lattice 
  as described in the main text. A rigid side chain (grey) is attached 
  to the fully flexible end-grafted backbone at monomer $b$ ('branching site'). 
  The substrate is impenetrable, i.e. the chain conformations are restricted 
  to the half-space $z>0$. Note that a chain of length $N$ consists of $N+1$ 
  monomers.} 
\label{fig:fig01}
\end{center}
\end{figure}

To simulate a flexible end-grafted polymer chain of length $N$ (i.e. having
$N+1$ monomers), we utilize a simple cubic lattice with unity lattice spacing (which
we have taken as our unit of length). Each monomer occupies a single lattice
site and self-avoidance is guaranteed by prohibiting multiple occupations. The
polymer is considered in a good athermal solvent, i.e. no interactions between
the monomers besides the self-avoidance are taken into account. The monomer with 
index~0 is placed in the plane $z=0$, and the halfspace $z<0$ is chosen as impenetrable 
while the lattice is taken large enough to avoid influences of any other boundary. A 
rigid side chain of length $S$ is attached at monomer $b$ of the flexible chain (counted 
from the grafted end). Figure~\ref{fig:fig01} shows a two-dimensional sketch of the model.

The flexible chain is simulated using a Verdier-Stockmayer-type algorithm
allowing for kink and 90$^{\circ}$/180$^{\circ}$-crank shaft moves 
(see, e.g.,~\cite{Binder1995}). Due to its rigidity, the only possible moves 
of the side chain are rotations by $\pm90^{\circ}$ and $180^{\circ}$ in the 
$xy$-, $yz$-, or $xz$-plane respectively around the branching site~$b$.
A side chain rotation can be induced by a regular move (kink, crank) involving
monomer $b$ or randomly without translocating monomer $b$, i.e. by introducing 
a new type of move. In every attempt only those moves are accepted that lead 
to self-avoiding conformations while respecting the impenetrability of the
substrate. Yet, 'during' a rotation process of the side chain the self-avoidance
can be violated. In connection with the non-ergodicity of all $N$-conserving
algorithms for SAWs~\cite{Madras1987}, this may lead to trapped conformations 
in the described model. To avoid an overrepresentation of trapped conformations, 
the simulation is continued with a newly generated random conformation when the 
system appears to have run into a dead end. In practice, this case occurred for 
rather short chains only.


\section{Results and Discussion}
To investigate the behavior of the described polymer system, we vary 
the lengths of the flexible backbone ($N$) and the rigid side chain ($S$) 
as well as the branching site~$b$ at which the side chain is attached to 
the backbone. As basic read-outs, we then monitor the squared radius 
of gyration $R_g^2$, the principle moments of gyration and the angle $\phi$ 
between the longest axis of the gyration ellipsoid and the surface normal, 
defined by $\cos\phi=\mathbf v_3\cdot\mathbf e_z/L_3$ (cf. \Eq{gyrtensor}).
All these quantities are calculated from the backbone monomers only, in 
order to study the conformational differences of an end-grafted polymer with
an attached side chain to one without.
\begin{figure} 
  \includegraphics[width=7cm]{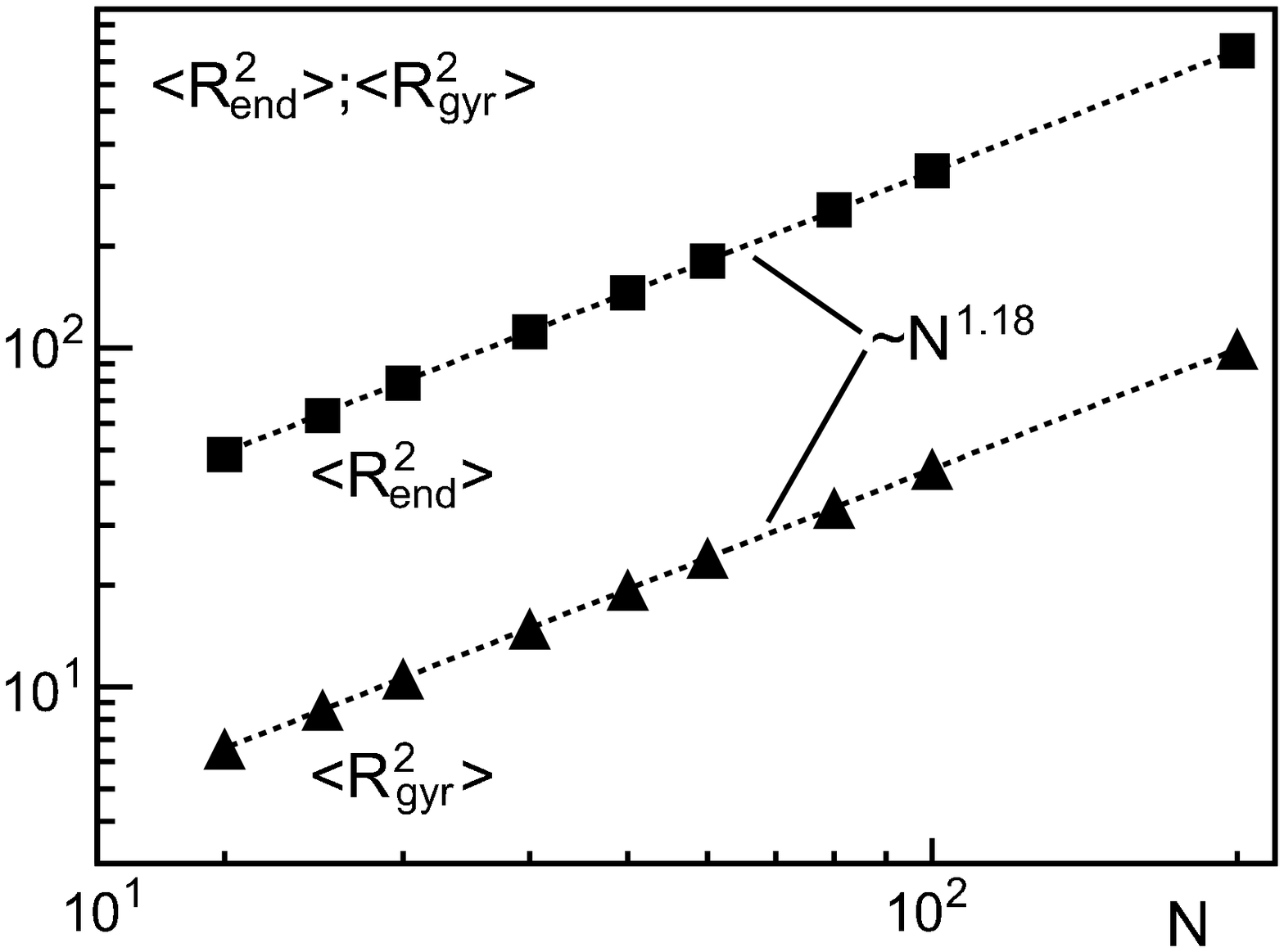}\\
  \includegraphics[width=7cm]{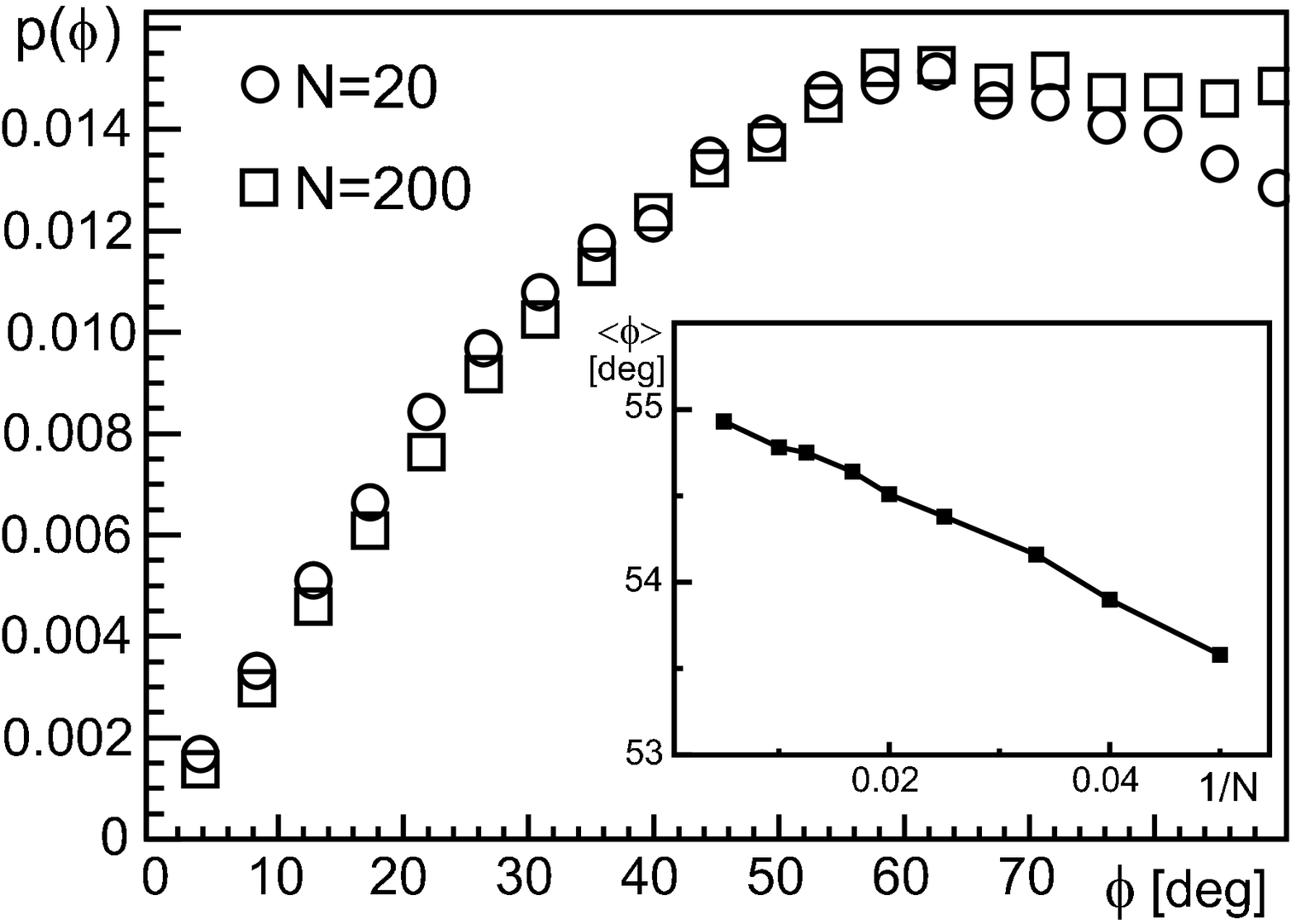}
  \caption{(a) The radius of gyration increases like $R_0^2\sim N^{2\nu}$
      ($\nu\approx 0.59$) for a flexible end-grafted chain (without side chain). 
      (b) The probability density function $p(\phi)$ of the orientation angle of the
      polymer's gyration ellipsoid shows a plateau-like behavior for $\phi \to 90^{\circ}$
      in the limit of large $N$. Inset: The average angle $\langle\phi\rangle$,
      i.e. the first moment of $p(\phi)$, depends only weakly on $N$ with 
      an asymptotic value $\langle \phi \rangle=55^{\circ}$.}   
\label{fig:fig02}
\end{figure}

We first determine the squared radius of gyration $R_0^2$ and the orientation 
angle $\phi$ of the gyration ellipsoid for an end-grafted flexible backbone 
($N = 20, \dots, 200$) {\em without} side chain. In agreement with~\cite{Huang2001}, 
we observe an increase $R_0^2\sim N^{2\nu}$ with $\nu\approx 0.59$ (\fig{fig:fig02}a). 
In contrast, the average angle of orientation, $\langle\phi\rangle$, only depends 
very weakly on $N$ and tends towards an asymptotic value $\langle\phi\rangle=55^{\circ}$ 
(\fig{fig:fig02}b, inset). The entire probability density function $p(\phi)$ of the 
orientation angles for $N=20$ and $N=200$ is shown in \fig{fig:fig02}b. Clearly, in 
both cases the mean $\langle\phi\rangle$ does not coincide with the plateau of the 
most probable value of $\phi$ near to 90$^{\circ}$, i.e. the gyration ellipsoid is most 
likely oriented parallel to the substrate ('mushroom'). It is noteworthy, that the 
plateau-like behavior of $p(\phi)$ only emerges properly for large $N$, while a 
too short polymer shows a slight decrease of $p(\phi)$ for $\phi \to 90^{\circ}$.

\begin{figure} 
  \includegraphics[width=7cm]{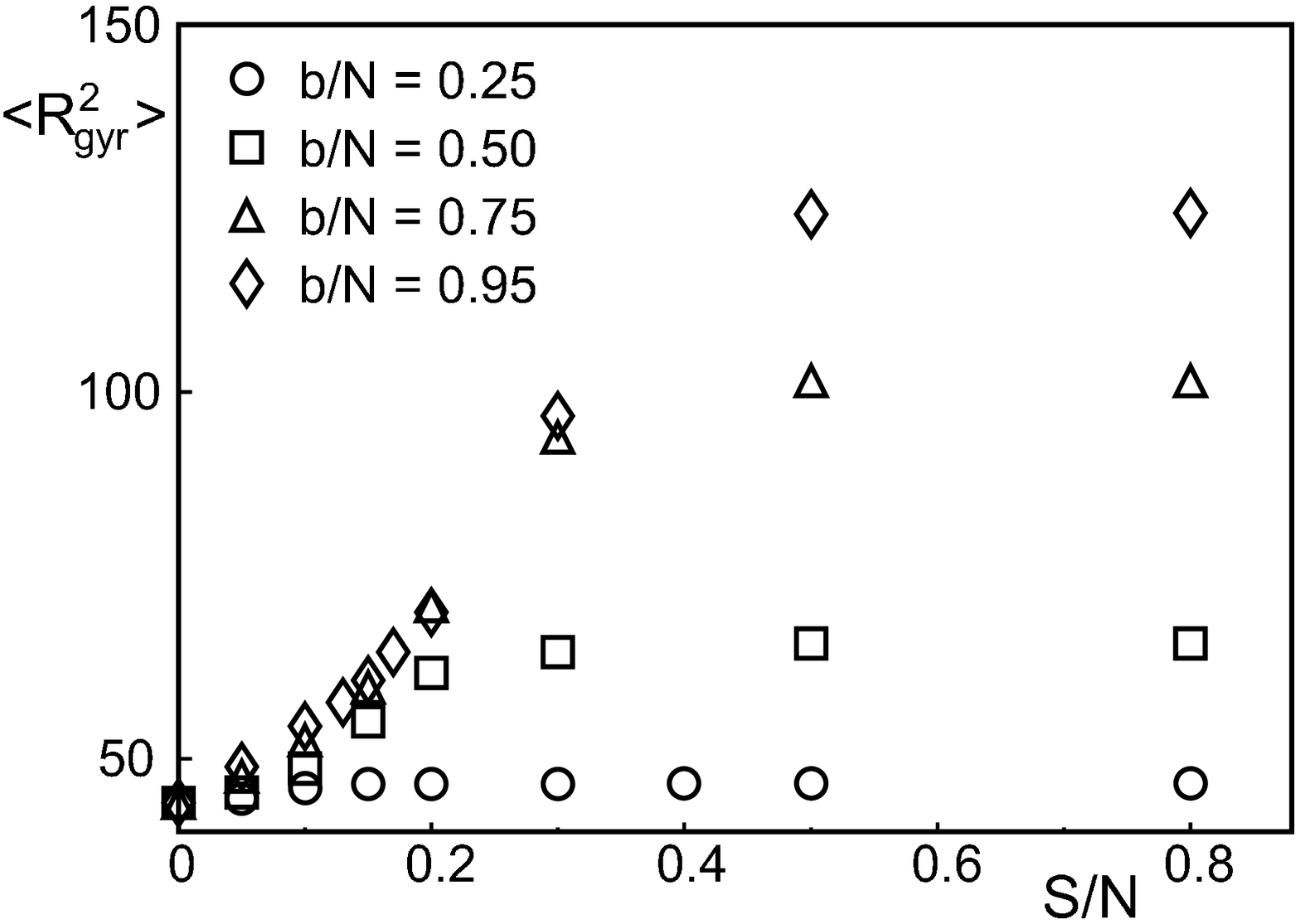}\\
  \includegraphics[width=7cm]{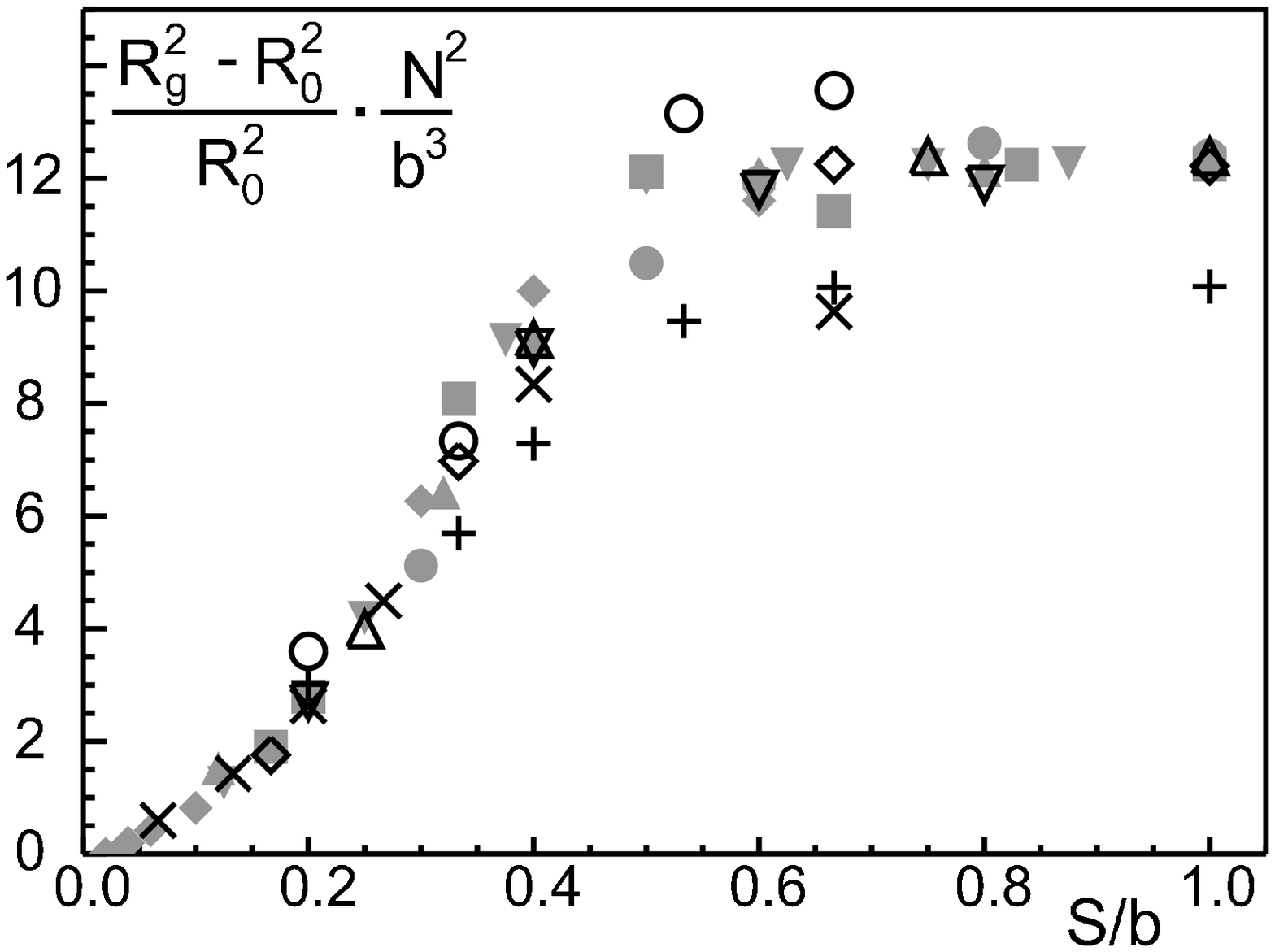}
  \caption{(a) The squared radius of gyration $\langle R_g^2 \rangle$ increases sigmoidally with 
      the length of the side chain $S/N$ and levels off at increasing values
      with increasing $b$ (here: $N=100$). (b) All curves collapse to a single master curve
      when applying the scaling \Eq{Rg_scale}. Open black symbols: $b/N=0.25$,
      $N=20,60,80,100,120$; filled grey symbols: $b/N=0.5$, $N=20,50,80,100,120$; 
      crosses: $b/N=0.75$, $N=20,100$.}
\label{fig:fig03}
\end{figure}

When attaching a rigid side chain of length~$S$ at position~$b$ to the
flexible backbone (length~$N$), we observe an increase in the radius of 
gyration that depends on $S$, $b$ and $N$. A representative example ($N=100$)
is shown in \fig{fig:fig03}a. The squared radius of gyration increases
sigmoidally from the unperturbed value $R_0^2$ with increasing $S$ whereas 
the limiting plateau strongly depends on $b$. The gross shape of the curves 
can be rationalized by considering the limiting cases: for $S\to0$ and 
$b\to 0$ the unperturbed backbone has to be recovered as the side chain 
vanishes or appears to be 'glued' to the substrate surface, respectively. 
For large $S$ any further increase of the side chain is not 'felt' by the 
mushroom-like backbone, i.e. the curve should level off to a constant. In
fact, we are able to collapse all data for the side-chain induced 
difference $\Delta =R_g^2-R_0^2$ to a single master curve (\fig{fig:fig03}b) 
for various combinations of $N$ and $b$ by the scaling:
\begin{equation}\label{Rg_scale}
  x=\frac{S}{b}\qquad y=\frac{R_g^2-R_0^2}{R_0^2}\cdot\frac{N^2}{b^3}
\end{equation}
As a result of \Eq{Rg_scale}, we clearly observe that $\Delta \sim R_0^2b^3/N^2$ 
in the plateau region. Thus, for $b=N$ (side chain attached to the free end) 
the gain in length follows approximately $\Delta \sim N^2$ which is reminiscent of 
the behavior of flexible end-grafted chains in the 'regime of stretched chains' 
observed in polymer brushes~\cite{Gennes1980}. 

\begin{figure} 
  \includegraphics[width=7cm]{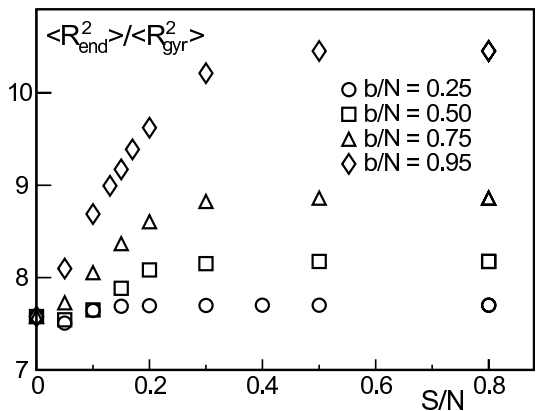}
  \caption{The ratio $\theta=\langle R_{\rm end}^2 \rangle/\langle R_g^2 \rangle$ of the squared 
  radius of gyration and the squared end-to-end vector of the backbone increases with increasing
  length of the rigid side chain $S/N$ (here: $N=100$).}
\label{fig:fig04}
\end{figure}

In \fig{fig:fig04}, we have plotted the ratio between the mean squared end-to-end distance and the
mean squared radius of gyration, $\theta=\langle R_{\rm end}^2\rangle/\langle R_g^2 \rangle$, 
as a function of~$S$. For a simple, free self-avoiding walk this ratio can be calculated to 
be~$\theta=6$ in the limit $N \to \infty$ while $\theta\approx 7.6$ for a chain attached to a 
solid substrate (see offset in \fig{fig:fig04} for $S=0$). Attaching a side chain results in 
a further increase of $\theta$, indicating that $\langle R_{\rm end}^2 \rangle$ grows faster 
than $\langle R_g^2 \rangle$. Again, the effect is strongest for large values of $S$ and $b$. 
This observation yields further evidence that the side chain, due to imposing an excluded-volume 
constraint, causes the felxible backbone to swell and disentangle to a more straight conformation. 
This picture is confirmed and refined by the subsequent results obtained from the study of 
the gyration tensor.
%
%

While $\langle R_g^2 \rangle$ measures the spherical size of a polymer its actual shape is 
better described by the dimensions of the gyration ellipsoid $L_1,\,L_2,\,L_3$. The ratio 
$\langle L_1^2 \rangle:\langle L_2^2 \rangle: \langle L_3^2 \rangle$ resembles the asphericity of 
the polymer. Our results show that for growing $S$ $\langle L_3^2 \rangle : \langle L_1^2 \rangle$ 
increases up to a plateau at $S \approx 0.5 N$, that depends on $b$ -- analogous to 
$\langle R_g^2 \rangle$. For $N=100$ and $b=0.5 N$ we obtain a maximum ratio of approximately 
$26:1$, indicating a dramatically altered conformation in comparison to a simple end-grafted chain, 
where $14.9:1$ is found~\cite{Huang2001}. The ratio is even higher for larger values of $b$, that is,
the backbone takes on a pronounced rod-like shape due to the attached rigid side chain.

The ratio $\langle L_2^2 \rangle : \langle L_1^2 \rangle$ between the two shorter axes of gyration shows 
a different behavior: For $b \lessgtr 0.5N$ it changes only marginally with increasing $S$ and takes on
a value of about $3.1:1$ independent of $b$. In contrast, for $b=0.5N$ a small but appreciable increase 
occurs leading to a saturation value of $4:1$ for $N=100$.

\begin{figure} 
  \includegraphics[width=7cm]{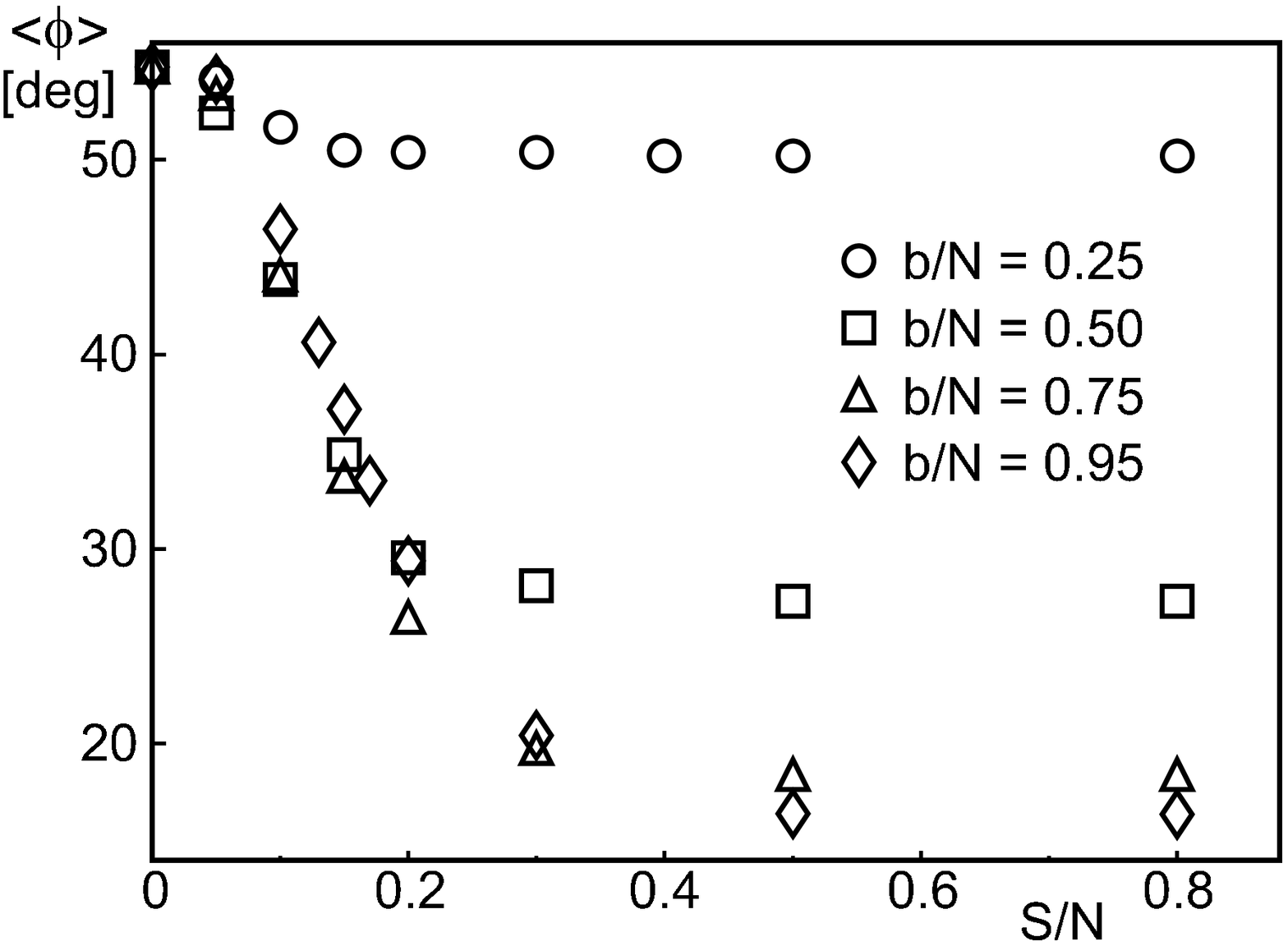}\\
  \includegraphics[width=7cm]{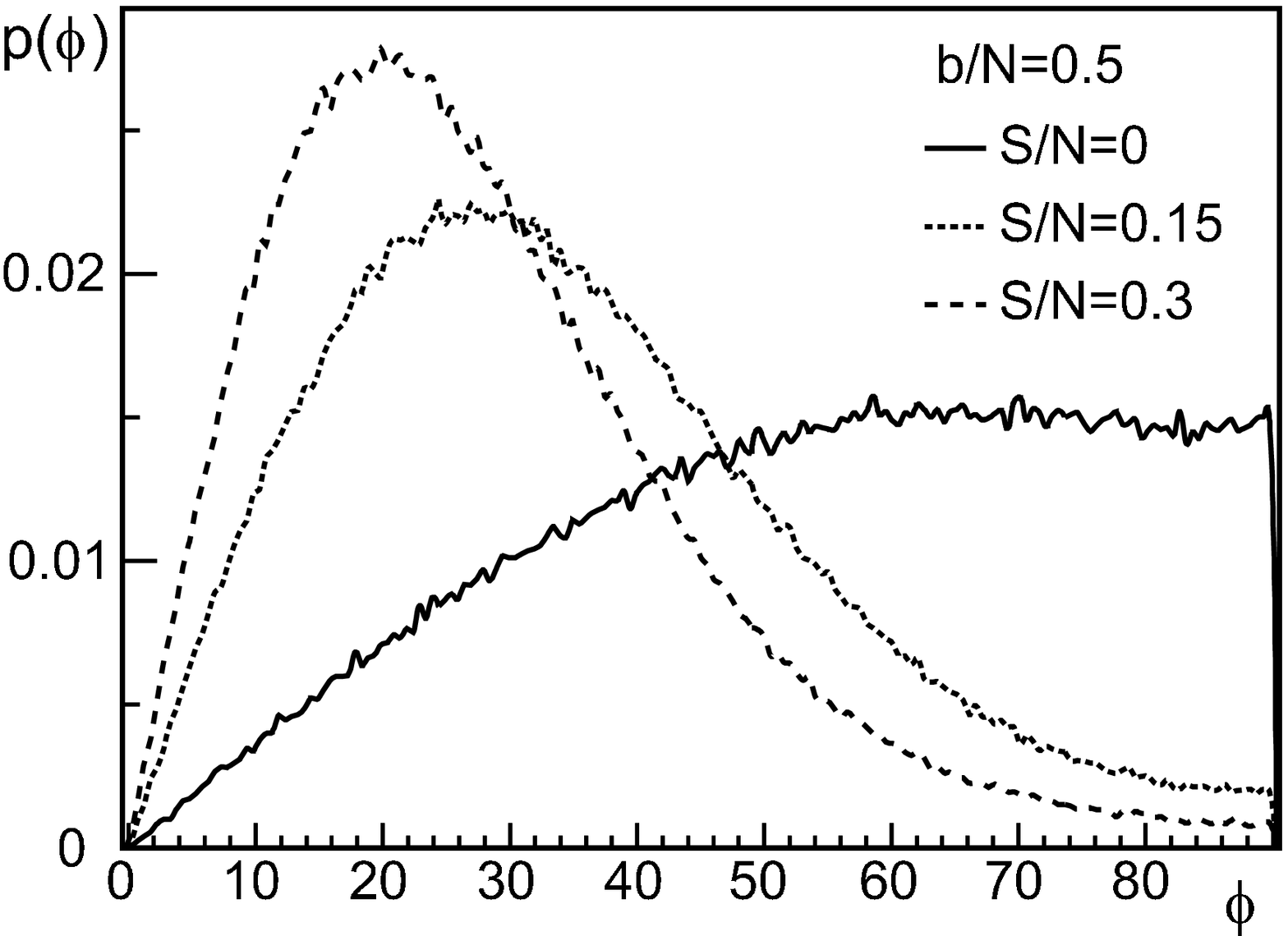}
  \caption{(a) The average angle $\langle\phi\rangle$ between the longest axis of 
  gyration and the substrate normal decreases for increasing lengths of the side chain 
  $S/N$, indicating a straightening of the backbone. Moving the branching site $b$ towards 
  the free end of the backbone strongly enhances this straightening. (b) The entire
  distribution $p(\phi)$ shifts and becomes more narrow as $S/N$ is decreased
  (here: $N=100$).}
\label{fig:fig05}
\end{figure}

We next consider the orientation of the backbone in terms of the the distribution of orientation 
angles $p(\phi)$ between the longest principle axis of gyration, $\mathbf v_3$, and the surface 
normal. As can be seen in \fig{fig:fig05}a for the representative example $N=100$, elongating 
the side chain leads to a considerable decrease of the average angle $\langle\phi\rangle$, indicateing
a more brush-like configuration of the backbone. This straightening becomes more pronounced when the 
branching site $b$ is moved towards the free end of the backbone, i.e. for large $S$ and $b\to N$ 
the backbone deviates from the surface normal by less than $20^\circ$. Concomitant to the decrease 
of the average orientation angle, the entire distribution $p(\phi)$ changes and assumes a more 
compact, i.e. narrow, shape around the mean $\langle\phi\rangle$ for large $S$ (\fig{fig:fig05}b). 
The decrease in width of the distribution accompanying the decrease of the average highlights 
the brush-like conformation of the backbone.
%

\begin{figure} 
  \includegraphics[width=7cm]{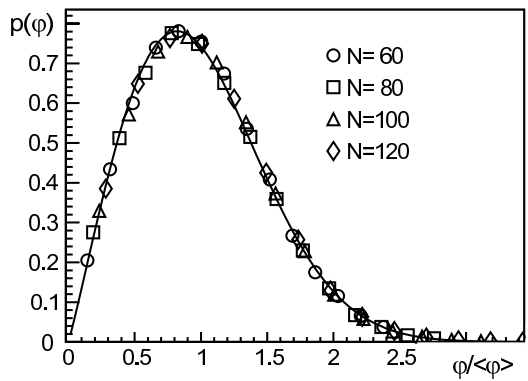}
  \caption{Normalizing the average angle, i.e. setting
  $\varphi=\phi/\langle\phi\rangle$, for $b\approx N$ leads to a collapse of
  all distributions $p(\varphi)$ for various $N,S$ with $S/N=0.3$ (symbols). The 
  curve is best described by \Eq{phi_pdf} (full line).
}
\label{fig:fig06}
\end{figure}

Indeed, the compact shape of $p(\phi)$ is a generic feature of the backbone for $b\to N$ and 
large $S$. This is reflected by the fact that all distributions can be collapsed to a single 
curve master curve 
\begin{equation}\label{phi_pdf}
  p(\varphi)=\frac{k\varphi^{k-1}}{\lambda^k}\exp\left\{-\left(\frac{\varphi}{\lambda}\right)^k\right\}
\end{equation}
with $k\approx2,\,\lambda\approx1$, when considering the normalized angle $\varphi=\phi/\langle\phi\rangle$ 
and fixing $b/N=const$ and $S/N=const$ (\fig{fig:fig06}). 

\section{Conclusions}
We conducted Monte Carlo simulations using a flexible polymer backbone, end-grafted 
to a solid substrate, with a rigid side chain attached to it. From the behavior 
of the backbone's radius of gyration and the length and orientation of its
longest principle axis of gyration, we are able to conclude that attaching a rigid 
side chain leads to a straightening of the backbone to a more brush-like configuration.
Depending on the side chain length $S$ and the branching site $b$, the radius of gyration 
and the gyration component perpendicular to the substrate is enhanced while the average 
orientation tends towards the surface normal. The effects of the side chain are strongest 
in the case of large $b$ and $S$. 

For an undisturbed end-grafted chain (\fig{fig:fig02}b) the probability density $p(\phi)$ 
shows a plateau for angles near to 90$^{\circ}$, resembling a coiled conformation 
(mushroom). Attaching a stiff side chain leads to a shift to smaller angles and 
a narrowing of $p(\phi$) (\fig{fig:fig05}b). It is found that $p(\phi)$ can be 
scaled using a Weibull distribution. According to the changes of $p(\phi)$ the arithmetic 
mean $\langle \phi \rangle$ changes starting at about 55$^{\circ}$ in the case of 
an undisturbed chain to smaller values when a side chain is attached (\fig{fig:fig05}a).

The behavior of the radius of gyration as well as of the ratio 
$\langle R_g^2 \rangle/\langle R_{\rm end}^2 \rangle$ indicates a swelling and 
straightening of the backbone due to the presence of the stiff side chain.
The backbone's change from a mushroom to a more rod-like conformation is  
also reflected in the ratios between the axes of gyration: While the two smaller
principal axes only show minor changes, the longest principal axis is
increased manifold, e.g. 26-fold for $b/N=1/2$. 
Furthermore, the increase in the radius of gyration follows a heuristic scaling 
with maximum value $b^3/N^2$ which implicates a more brush-like growth of $R_g^2$ with 
the backbone length $N$. 

It will be interesting to examine the influence of a semi-flexibility of the side chain
to obtain a more realistic representation of the above described HA-aggrecan system.  
To approach the biological setting, e.g. the protective extracellular matrix, the 
phase diagram of end-grafted polymers with an attached side chain at varying surface densities 
will be of great interest. Work along these lines is currently underway.

\begin{acknowledgments}
We would like to thank Dennis Gro\ss e for providing the initial code and Manfred Bohn for 
helpful discussions. This work was supported by the Institute for Modeling and Simulation in
the Biosciences (BIOMS) in Heidelberg.
\end{acknowledgments}


\end{document}